\documentclass[aps,orb,twocolumn,amssymb,showpacs,pre]{revtex4}
\usepackage{graphicx}
\usepackage[dvipdf]{epsfig}
\usepackage{subfigure}
\usepackage{color}
\usepackage{amstext}
\usepackage{amsmath}
\usepackage{amssymb}
\usepackage{graphicx}
\usepackage{dcolumn}
\usepackage{bm}
\usepackage{rotating}

\begin{document}
\def\be{\begin{equation}}
\def\ee#1{\label{#1}\end{equation}}
\def\no{\nonumber}
\def\lb{\label}
\newcommand{\ben}{\begin{eqnarray}}
\newcommand{\een}{\end{eqnarray}}
\def\bx{\mathbf{x}}
\def\bc{\mathbf{c}}
\def\bv{\mathbf{v}}
\def\bp{\mathbf{p}}
\def\bk{\mathbf{k}}
\def\k{\textsf{k} }
\def\z{\zeta}

\title{Self-gravitating systems of ideal gases in the 1PN approximation}
\author{Gilberto M. Kremer$^1$}\email{kremer@fisica.ufpr.br}\author{Mart\'in G. Richarte$^{1,2}$}\email{martin@df.uba.ar}
\author{Karla Weber$^1$}\email{karlaw@fisica.ufpr.br}
\affiliation{$^1$Departamento de F\'isica, Universidade Federal do Paran\'a, Caixa Postal 19044, 81531-990 Curitiba, Brazil}
\affiliation{$^2$Departamento de F\'isica, Facultad de Ciencias Exactas y Naturales,
Universidad de Buenos Aires, Ciudad Universitaria 1428, Pabell\'on I,  Buenos Aires, Argentina}

\begin{abstract}
We obtain the Maxwell-J\"uttner distribution function  at first order in the post-Newtonian approximation within the framework of general relativity.
Taking into account the aforesaid distribution function, we compute the particle four-flow and energy-momentum tensor.
We focus on the search of static solutions for the gravitational potentials with spherical symmetry. In doing so, we obtain the density, pressure  and gravitational potential energy profiles in terms of dimensionless radial coordinate by solving the aforesaid equations numerically.  In particular, we find the parametric profile for the equation of state $p/\rho$ in terms of the dimensionless radial coordinate. Due to its physical relevance,  we also find the galaxy rotation curves using  the post-Newtonian approximation. We join two different kinds of static solutions in order to account for  the linear regime near the center  and the typical flatten behavior at large radii as well.
\end{abstract}
 \maketitle

\section{Introduction}

Galaxies are composed of many different kinds of astrophysical objects such  as stars, interstellar gas, dust, and dark matter amongst other things. Such system  holds together due to gravitational interaction amongst their constituents.  In fact, they can be organized in bigger astrophysical structures called clusters or super-clusters \cite{ga}. It is generally believed that the amount of dark matter in a cluster is ten times bigger than the total amount of gas and stars \cite{liddle}. However, the recent analysis based on both optical and radio data in spiral galaxies reveals that  the amount of dark matter is even less than previously thought \cite{jaloncha};  where the gravitational field  associated with spiral galaxies (such as NGC 7793, 1365, 6946 and UGC 6446) is assumed to follow a global disk-like profile rather than spheroidal one \cite{jaloncha}. For a detailed analysis of the universal rotational curve of spiral galaxies and its connection with dark matter see \cite{paolo}.
If the physical scale of interest is much bigger than the size of  these objects then the latter ones can be considered as pointlike particles interacting among them,  the gravity being the most relevant interaction. Hence, one possible route to study this system may be  to solve its dynamical equation and find the trajectories of each one of the ``particles''.  Nevertheless, the number of particles involved in these systems is enormously big so the aforesaid approach becomes unsuitable. An alternate view of galaxies is as a system of particles in
  six dimensional phase space. The galaxies are then instantaneously described in terms of a distribution function $f (x , p )$ over the phase space, where $x$ represents the position vector and $p$ stands for the momentum. In this way, galaxies can be described in terms of ensemble of particles  along with a distribution function which satisfies a kinetic equation \cite{b2}.  If one obtains  the distribution function   associated with the kinetic equation then one can extract the main traits of the system under consideration, for instance, one can be able to reconstruct the average square velocity and several other moments of the distribution function\cite{b2}.

If the collision between two particles is  a very rare event then the collision operator of the Boltzmann equation can be neglected. In the latter case, one can  work with the collisionless Boltzmann equation (sometimes this is also known as Vlasov equation)   whose solution turns  out to be  an equilibrium distribution function \cite{b2}.  As an example, one thinks in a gas with a Boltzmann distribution at equilibrium whose gravity center follows a circular geodesic in Schwarzschild field \cite{ZK}. Or the case of  dark matter halos,   dark matter only interacts gravitationally and there are hardly any encounters, so that one can describe the dark matter as a collisionless system and use the Vlasov equation in order to describe its evolution \cite{zemp}. Besides, a self-consistent rederivation of collisionless  Boltzmann equation
for self-gravitating gases  with post-Newtonian  corrections \cite{b1} was obtained recently; the case of post-Newtonian polytropes solutions was examined numerically, focusing in the role played by  the relativistic correction to the rotation curves \cite{b1}.

As is well-known the equations of general relativity  reduce to those of Newtonian gravity in the limit of slow motions
along with  weak gravitational fields. Newton's theory of gravity is good enough to describe all the physics in the solar system, but  it is 
incomplete and requires some corrections to properly account for the shift in the perihelion of Mercury \cite {Wein}. In order to describe
physical phenomena like the latter one it becomes essential to include post-Newtonian (PN) corrections to the standard gravitational physics.
A straightforward manner to include all the relativistic corrections  to Newtonian gravity is by means of post-Newtonian formalism.
The post-Newtonian method relies on the idea that one perturbs around a Minkowski background. In doing so,  one considers that the perturbations of the metric tensor along with those corresponding to  the energy-momentum tensor are both small fluctuations, which can be expanded in power of $v/c$, being $v$ a typical velocity  in a system associated with matter and $c$ is the speed of light \cite {Wein}.   One of the main reasons to  examine relativistic corrections in galactic dynamics concerns to the issue of
the rotation curves predicted by Newtonian theory. In Newtonian theory, rotation curves typically increase linearly near the origin up to a maximum and then vanish for large radii, while the measured circular velocity
curve for the galaxies leads to small value near the center,
increases linearly, then exhibits a small cusp and tends to a
finite nonzero value at large radii.  One can clearly recognize the contribution of the dark matter halo in order to generate the flat circular velocity curve \cite{zemp}. Indeed, some authors pointed out that the inclusion of post-Newtonian corrections in astrophysical models can really help to reduce the amount of dark matter needed to explain rotation curves which flatten at large radii \cite{grumiller}. Of course,  these corrections by their own cannot overcome the whole problem of generating flat rotation curves, however, can reduce the amount of dark matter in relation with the Newtonian models.

For the reasons mentioned above, the study of  relativistic corrections within the context of kinetic theory seems to be a very promising route. Bearing this in mind,  many authors  devoted some efforts to apply kinetic theory  plus post-Newtonian method to describe astrophysical models. They explored the kinetic theory of self-gravitating gases with axial symmetry within the framework of post-Newtonian formalism applied to the case of razor thin disks, focusing on axially symmetric galaxy models\cite{b1a}. In particular, they obtained the rotation curves and mass profiles for the post-Newtonian version of Morgan-Morgan disks and derived  the virial theorem in  1PN approximation as well \cite{b1a}. Continuing with this line of research, Nguyen and Pedraza studied self-gravitating system  with polytropic equation of state in the PN approximation of general relativity, the physical motivation is that such equation can be used to describe  many astrophysical models such as  white dwarfs, neutron stars, galactic halos, and globular clusters amongst others.  By solving the Einstein-Vlasov system of equations, they found a family of star clusters with anisotropic in velocity space within PN scheme. In addition, they analyzed the stability of circular orbits for radial perturbations \cite{b1b}.

Here, we are going to use the relativistic collisionless Boltzmann equation along with   the Maxwell-J\"uttner distribution function to derive the
Maxwell-J\"uttner distribution function in the 1PN approximation. In doing so, we introduce a peculiar velocity associated with the particle velocity in the gas frame, use the Tolman's law and integrate over the  peculiar velocity space.  We calculate the particle flow  at 1PN order while energy--momentum tensor components are derived at different PN orders, for instance, the diagonal temporal part is calculated at 1PN order but the diagonal spatial components are reported at 2PN order. In this way, we derive the macroscopic energy momentum tensor from the Maxwell-J\"uttner distribution function at  1PN order. Our result coincides with the one obtained from fluids description \cite{Wein}. We numerically solve the nonlinear equation associated to  the gravitational potential fields. Our analysis differs from the one reported for post-Newtonian polytropes in diverse manners \cite{b1}. We contrast the Newtonian with the PN profiles of matter density, circular velocity and gravitational potential energy. As  was expected, the pressure profile only includes PN corrections. For certain value radius, say $r_c$,  the gravitational fields become complex and therefore we must match the latter solution with a physical one. We show the procedure to obtain physical gravitational potentials by joining with other solutions at $r_c$. In addition, we notice that such method leads to circular velocity profiles that flatten at large radii.

Our paper is outlined as follows. In Sec. II, we give the general derivation of the Maxwell-J\"uttner distribution function at  1PN order. We devote Sec. III to obtain the particle four-flow and energy-momentum tensor within the PN approximation from the Maxwell-J\"uttner distribution function at  1PN order. We seek static solution for the gravitational potentials associated with non-linear Poisson-like equations in Sec. IV. In Sec. V, we solve numerically the aforesaid equations, present the  density, pressure and circular velocity profiles. In Sec. VI, we finally reexamine the issue of generating flat rotation curves by gluing two different solutions at certain radius.  In Sec. VII the conclusions are stated. Throughout the article we   adopt  the metric signature $(-,+,+,+)$ and we do not set  the speed of light $c$  equal to the unity for practical reasons. 
\section{The Maxwell-J\"uttner distribution function in the 1PN approximation}

The kinetic theory of relativistic gases in the presence of gravitational fields is based in the Boltzmann equation (see e.g.\cite{b2} and \cite{b3}-\cite{b6})
\ben\lb{1}
p^{\mu}\frac{\partial f}{\partial x^{\mu}}-\Gamma_{\mu\nu}^{i}p^{\mu}p^{\nu}\frac{\partial f}{\partial p^{i}}=\mathcal{C}(f,f).
\een
Here $f\equiv f(\bx,\bp,t)$ is the one-particle distribution function in the phase space spanned by the space $\bx$ and three-momentum $\bp$ coordinates,  $\Gamma_{\mu\nu}^{i}$ are  Christoffel symbols and $\mathcal{C}(f,f)$ is the collision operator of  the Boltzmann equation which takes into account the product of the two distribution functions of the colliding particles. In the above expression the mass-shell condition $p^\mu p_{\mu}=-m^2c^2$ -- where $m$ is the particle rest mass  -- was taken into account.

Equilibrium is characterized by a vanishing collision term $\mathcal{C}(f,f)$ and this implies that the distribution function is the Maxwell-J\"uttner distribution function
\ben\lb{2}
f=\frac{n}{4\pi m^2ckTK_2(\zeta)}\exp\left(\frac{p^\mu U_\mu}{kT}\right).
\een
Above $k$ is the Boltzmann constant and $n,T, U^\mu$ are the particle number density, the  temperature and the   four-velocity of the gas, respectively. Furthermore, $\zeta=mc^2/kT$ represents the ratio of the rest energy of a particle $mc^2$ and the thermal energy of the gas $kT$ and $K_2(\z)$ denotes the modified Bessel function of second kind.  The ultrarelativistic regime of the gas is attained at high temperatures where $\z\ll1$, while the nonrelativistic one occurs at low temperatures where $\z\gg1$. Another expression for the Maxwell-J\"uttner distribution function is given in terms of the chemical potential $\mu$ -- the Gibbs function per particle -- namely,
\ben\lb{3}
f=\exp\left(\frac\mu{kT}-1+\frac{p^\mu U_\mu}{kT}\right).
\een

If we insert the equilibrium distribution function into the left-hand side of the Boltzmann equation (\ref{1}) we obtain the following restrictions for the temperature and chemical potential fields (see \cite{ZK})
\ben\lb{4a}
T\sqrt{-g_{00}}=T_0={\rm constant},\\\lb{4b}
\mu\sqrt{-g_{00}}=\mu_0={\rm constant}.
\een
The first equation above is known as Tolman's law \cite{To1,To2} while the second one as Klein's law \cite{Kl} and both were introduced within the framework of phenomenological theories.

From the two expressions for the Maxwell-J\"uttner distribution function (\ref{2}) and (\ref{3})  we get
\ben\lb{5}
\frac{n}{4\pi m^2ckTK_2(\zeta)}=\exp\left(\frac\mu{kT}-1\right)
=\exp\left(\frac{\mu_0}{kT_0}-1\right),\quad
\een
where the last equality follows from Tolman (\ref{4a}) and Klein (\ref{4b}) laws. Hence we may write the left-hand side of (\ref{5}) in terms of $n_0$ and $T_0$, namely,
\ben\lb{6}
\frac{n}{4\pi m^2ckTK_2(\zeta)}=\frac{n_0}{4\pi m^2ckT_0K_2(\zeta_0)}.
\een

Let us calculate the Maxwell-J\"uttner distribution function (\ref{2}) in  the first post-Newtonian approximation (1PN approximation).

First we follow \cite{Wein} and write the line element
\ben\lb{7a}
ds^2=-c^2 d\tau^2=g_{00}(dx^0)^2+2g_{0i}dx^0dx^i+g_{ij}dx^idx^j,\qquad
\een
up to $1/c^4$  order as
\ben\no
c^2 d\tau^2=\left(1+\frac{2\phi}{c^2}+\frac{2(\phi^2+\psi)}{c^4}\right)(dx^0)^2
\\\lb{7b}
-2\frac{\xi_i}{c^3}dx^0dx^i-\left(1-\frac{2\phi}{c^2}\right)\delta_{ij}dx^idx^j,
\een
thanks to the expressions  of the metric tensor components (\ref{a1}) -- (\ref{a5}) given in the Appendix A.
Next by introducing the four-velocity of the gas particles $u^\mu=(u^0,u^0v^i/c)$ it follows from (\ref{7b}) that up to $1/c^4$  order we have
\ben\no
\frac{u^0}c&=&1+\frac1{c^2}\left(\frac{v^2}2-\phi\right)
\\\lb{8}
&+&\frac1{c^4}\left(\frac{3v^4}8-\frac{5v^2\phi}2+\frac{\phi^2}2-\psi+\xi_iv^i\right).
\een

The above expression is also valid for the components of the fluid four-velocity  $U^\mu=(U^0,U^0V^i/c)$ by replacing $u^0$ and $v^i$ by $U^0$ and $V^i$, respectively.

If we introduce the peculiar velocity $\bf W=v-V$ -- which is the particle velocity in the gas frame -- and by considering Tolman's law (\ref{4a}) we get the following relationship
\ben\no
\frac{g_{\mu\nu}p^\mu U^\nu}{kT}=-\frac{mc^2}{kT_0}\left\{1+\frac{W^2}{2c^2}+\frac\phi{c^2}+\frac1{c^4}\left[\frac{3W^4}8\right.\right.
\\\no
\left.\left. -\frac{3W^2\phi}2+\frac {({\bf V\cdot W})^2}2+\frac{V^2W^2}2\right.\right.
\\\lb{9}
\left.\left.+({\bf V\cdot W})W^2+\frac{\phi^2}2+\psi\right]\right\}.
\een
Furthermore, up to  $1/c^2$ order the modified Bessel function of second kind reads
\ben\lb{10}
\frac1{K_2(\zeta_0)}=\sqrt{\frac{2mc^2}{\pi kT_0}}\,e^{\frac{mc^2}{kT_0}}
\left(1-\frac{15kT_0}{8mc^2}+\dots \right).
\een
Hence the Maxwell-J\"uttner distribution function (\ref{2}) in the 1PN approximation becomes
\ben\no
f=\frac{n_0}{(2\pi mkT_0)^\frac32}e^{-\frac{mW^2}{2kT_0}-\frac{m\phi}{kT_0}}
\bigg\{1-\frac{15kT_0}{8mc^2}
\\\no
-\frac{m}{kT_0c^2}\bigg[\frac{3W^4}8-\frac{3W^2\phi}2+\frac {({\bf V\cdot W})^2}2+\frac{V^2W^2}2
\\\lb{11}
+({\bf V\cdot W})W^2+\frac{\phi^2}2+\psi\bigg]\bigg\},\quad
\een
thanks to (\ref{8}), (\ref{9}) and (\ref{10}) and by considering the terms with the factor $1/c^2$ of small order.

\section{Macroscopic description}

In kinetic theory of relativistic gases the particle four-flow $N^\mu$ and  the energy-momentum tensor $T^{\mu\nu}$ are given in terms of the distribution function by (see \cite{b2})
\ben\lb{12a}
N^\mu=m^3c\int u^\mu  f\frac{\sqrt{-g}\,d^3 u}{-u_0},\\\lb{12b}
T^{\mu\nu}=m^4c\int u^\mu u^\nu f\frac{\sqrt{-g}\,d^3 u}{-u_0}.
\een
Note that we have written the energy-momentum tensor in terms of the particle four-velocities $u^\mu$ instead of the particle four-momentum $p^\mu=mu^\mu$. Here we follow \cite{b2,b3,b4,b5,b6} and considered that the element of integration is the  invariant $\sqrt{-g}d^3p/(-p_0)$ with the covariant component $p_0$, which is different from $\sqrt{-g}d^3p/(p^0)$ in the 1PN approximation, which was adopted by  \cite{b1}.

In the 1PN approximation the differential element $d^3u$ in terms of the peculiar velocity $\bf W$ up to the  $1/c^2$ order is given by
\be
 d^3u=\left\{1+\frac1{c^2}\left[\frac{5(V^2+2({\bf W\cdot V})+W^2)}{2}-3\phi\right]\right\}d^3W.
\ee{13}
Furthermore, up to the  $1/c^2$ order we can build the relation
\ben\no
&&-(g_{00}u^0+g_{0i}u^i)=-g_{00}u^0\left(1+\frac{g_{0i}u^i}{g_{00}u^0}\right)
\\\lb{14}
&&\approx u^0\left(1+2\frac{\phi}{c^2}\right),
\een
thanks to the expressions of the metric tensor given in the Appendix A. Next up to the  $1/c^2$ order we have the following relationship
\ben\no
\frac{\sqrt{-g}\, d^3 u}{-(g_{00}u^0+g_{0i}u^i)}=
\bigg\{1+\frac1{c^2}\bigg[\frac{5(V^2+2({\bf W\cdot V})+W^2)}{2}
\\\lb{15}
-7\phi\bigg]
\bigg\}\frac{d^3W}{u^0}.\qquad
\een

Once we know the distribution function in the 1PN approximation we can calculate the particle four-flow and the energy-momentum tensor in this approximation. The particle four-flow $N^{\mu}=\sum_n{\buildrel\!\!\!\! _{n} \over{N^{\mu}}}$ and the energy-momentum tensor $T^{\mu\nu}=\sum_n{\buildrel\!\!\!\! _{n} \over{T^{\mu\nu}}}$ are split in different orders of the ratio $(\overline{v}/c)^n$ (see \cite{Wein}) where $\overline v$ can be identified with the thermal velocity of a particle $\overline v=\sqrt{kT_0/m}$.

First from (\ref{12a}) together with (\ref{11}) and (\ref{15}) we get through integration of the resulting equation that the time component of the particle four-flow becomes
\ben\no
N^0=nU^0={\buildrel\!\!\!\! _{0} \over{N^0}}+{\buildrel\!\!\!\! _{2} \over{N^0}}=n_0ce^{-\frac{m\phi}{kT_0}}\qquad\qquad
\\\lb{16a}
\times\bigg\{1+\frac1{c^2}\bigg[\frac{V^2}2-\frac{5\phi}2-\frac{m}{kT_0}\bigg(\frac{\phi^2}2+\psi\bigg)\bigg]\bigg\}.
\een
From the above equation and from the expression for $U^0$ it is easy to obtain that the particle number density is given by
\ben\lb{16b}
n=n_0e^{-\frac{m\phi}{kT_0}}\left\{1-\frac1{c^2}\left[\frac{3\phi}2+\frac{m}{kT_0}\bigg(\frac{\phi^2}2+\psi\bigg)\right]\right\}.
\een
Hence we can write (\ref{16a}) as
\ben\lb{16c}
{\buildrel\!\!\!\! _{0} \over{N^0}}+{\buildrel\!\!\!\! _{2} \over{N^0}}=nc\left[1+\frac1{c^2}\left(\frac{V^2}2-\phi\right)\right].
\een
Following the same methodology the space components of the particle number density read
\be
{\buildrel\!\!\!\! _{1} \over{N^i}}+{\buildrel\!\!\!\! _{3} \over{N^i}}=\left({\buildrel\!\!\!\! _{0} \over{N^0}}+{\buildrel\!\!\!\! _{2} \over{N^0}}\right)\frac{V^i}c=nV^i\left[1+\frac1{c^2}\left(\frac{V^2}2-\phi\right)\right].
\ee{16d}

The components of the energy-momentum tensor (\ref{12b}) are obtained in the same manner, yielding
\ben\lb{17a}
{\buildrel\!\!\!\! _{0} \over{T^{00}}}+{\buildrel\!\!\!\! _{2} \over{T^{00}}}&=&\varepsilon\left[1+\frac1{c^2}\left(V^2-2\phi\right)\right],
\\\lb{17b}
{\buildrel\!\!\!\! _{1} \over{T^{0i}}}+{\buildrel\!\!\!\! _{3} \over{T^{0i}}}&=&\frac{V^i}c\left[\varepsilon+p+\frac\varepsilon{c^2}\left(V^2-2\phi\right)\right],
\\\no
{\buildrel\!\!\!\! _{2} \over{T^{ij}}}+{\buildrel\!\!\!\! _{4} \over{T^{ij}}}&=&p\left[1+\frac{2\phi}{c^2}\right]\delta^{ij}
\\\lb{17c}
&+&\left[\varepsilon+p+\frac\varepsilon{c^2}(V^2-2\phi)\right]\frac{V^iV^j}{c^2}.
\een
In order to derive the above expressions (\ref{17a}) -- (\ref{17c}), we have used several nontrivial identities related with the integration of different moments for a Gaussian distribution, see Appendix B for further details.
Here the energy density $\varepsilon=\rho c^2$ -- where $\rho$ is the mass density -- and the pressure $p$ are given by
\ben\lb{17d}
\varepsilon=\rho c^2=mnc^2\left(1+\frac{3kT}{2mc^2}\right),\qquad p=nkT.
\een
The expressions (\ref{17a}) -- (\ref{17c}) are the same as those which are obtained from the representation of the energy-momentum tensor
\ben\lb{17e}
T^{\mu\nu}=pg^{\mu\nu}+(\varepsilon+p)\frac{U^\mu U^\nu}{c^2},
\een
by considering  the metric tensor and the four-velocity in the PN approach up to the fourth order (cf. \cite{Wein}).

\section{The search for static solutions}

In this section we search for static solutions of Einstein's field equations, which in the 1PN approximation reduces to the following equations for the gravitational potentials (see e.g. \cite{Wein})
\ben\lb{19}
\nabla^2\phi=\frac{4\pi G}{c^2} \,{\buildrel\!\!\!\! _{0} \over{T^{00}}},
\qquad
\nabla^2\psi=4\pi G \left({\buildrel\!\!\!\! _{2} \over{T^{00}}}+{\buildrel\!\!\!\! _{2} \over{T^{ii}}}\right).\quad
\een

We follow \cite{b1} and integrate the components of the energy-momentum tensor (\ref{12b}) in the range $[0,v_e]$, where $v_e$ is the escape velocity. The escape velocity is calculated from the expression for the energy in the 1PN approximation \cite{b1}
\ben\lb{18}
E=m\left(\frac{v^2}2+\phi+\frac{3v^4}{8c^2}-\frac{3v^2\phi}{2c^2}+\frac{\phi^2}{2c^2}+\frac\psi{c^2}\right),
\een
 by considering that a gas particle attains its   maximum value at $E=0$,
so that the gas particle is unable to leave the distribution of matter. According to \cite{b1} to be consistent with the 1PN approximation the escape velocity must given by $v_e=\sqrt{-2\phi}$.
 Now by considering a vanishing gas velocity $\bf V=0$ we can  perform the integrations and get the following results for the components of the energy-momentum tensor in the different orders of the ratio $(\overline{v}/c)^n$:
\ben\lb{20a}
{\buildrel\!\!\!\! _{0} \over{T^{00}}}=-\rho_0 c^2\left[2\sqrt{-\frac{\phi_*}\pi}-e^{-\phi_*}{\rm erf}\left(\sqrt{-\phi_*}\right)\right],\qquad
\\\no
{\buildrel\!\!\!\! _{2} \over{T^{00}}}=-\frac{\rho_0 kT_0}m\bigg[\bigg(3-\frac{23}2\phi_*-10\phi_*^2-2\psi_*\bigg)\sqrt{-\frac{\phi_*}\pi}\qquad
\\\lb{20b}
-\bigg(\frac{3}2-\frac72\phi_*-\frac12\phi_*^2-\psi_*\bigg)e^{-\phi_*}{\rm erf}\left(\sqrt{-\phi_*}\right)\bigg],\qquad
\\\no
{\buildrel\!\!\!\! _{2} \over{T^{ii}}}=-\frac{\rho_0 kT_0}m\bigg[\left(6-4\phi_*\right)\sqrt{-\frac{\phi_*}\pi}\qquad
\\\lb{20c}
-3e^{-\phi_*}{\rm erf}\left(\sqrt{-\phi_*}\right)\bigg].\qquad
\een
In the above equations we have introduced the dimensionless quantities
\ben\lb{20d}
 \phi_*=\frac{m}{kT_0}\phi,
\qquad
\psi_*=\left(\frac{m}{kT_0}\right)^2\psi.
\een
Furthermore,
${\rm erf}\left(\sqrt{-\phi_*}\right)$ is the error function and ${\buildrel\!\!\!\! _{2} \over{T^{ii}}}$ refers to the trace of ${\buildrel\!\!\!\! _{2} \over{T^{ij}}}$. We  identify ${\buildrel\!\!\!\! _{0} \over{T^{00}}}$ with the energy density $\rho_0 c^2$, while ${\buildrel\!\!\!\! _{2} \over{T^{00}}}$ and ${\buildrel\!\!\!\! _{2} \over{T^{ii}}}$ with the pressure $\rho_0 kT_0/m$.

We insert (\ref{20a}) -- (\ref{20c}) into (\ref{19}) and get the following coupled system of equations for the gravitational potentials
\ben\lb{21a}
\nabla^2\phi_*=-k_J^2 \left[2\sqrt{-\frac{\phi_*}\pi}-e^{-\phi_*}{\rm erf}\left(\sqrt{-\phi_*}\right)\right],\qquad
\\\no
\nabla^2\psi_*=-k_J^2 \bigg[\bigg(9-\frac{31}2\phi_*-10\phi_*^2-2\psi_*\bigg)\sqrt{-\frac{\phi_*}\pi}\qquad
\\\lb{21b}
-\bigg(\frac{9}2-\frac72\phi_*-\frac12\phi_*^2-\psi_*\bigg)e^{-\phi_*}{\rm erf}\left(\sqrt{-\phi_*}\right)\bigg],\qquad
\een
where $k_J=\sqrt{4\pi G\rho_0}/{\overline v}$ can be identified as the Jeans wavenumber.

The gravitational potentials are only functions of the radial coordinate $r$ so that system of equations (\ref{21a}) and (\ref{21b}) in spherical coordinates can written as
\ben\lb{22a}
\frac1{\widetilde r^2}\frac{d}{d\widetilde r}\left(\widetilde r^2\frac{d\widetilde \phi}{d\widetilde r}\right)=
\left[2\sqrt{\frac{\widetilde\phi}\pi}-e^{\widetilde\phi}{\rm erf}\left(\sqrt{\widetilde\phi}\right)\right],\qquad
\\\no
\frac1{\widetilde r^2}\frac{d}{d\widetilde r}\left(\widetilde r^2\frac{d\widetilde \psi}{d\widetilde r}\right)=
\bigg[\bigg(9+\frac{31}2\widetilde\phi-10\widetilde\phi^2+2\widetilde\psi\bigg)\sqrt{\frac{\widetilde\phi}\pi}\qquad
\\\lb{22b}
-\bigg(\frac{9}2+\frac72\widetilde\phi-\frac12\widetilde\phi^2+\widetilde\psi\bigg)e^{\widetilde\phi}{\rm erf}\left(\sqrt{\widetilde\phi}\right)\bigg].\qquad
\een
Here the new dimensionless quantities are $\widetilde r=r k_J$, $\widetilde\phi=-\phi_*$ and $\widetilde\psi=-\psi_*$.

The system of equations (\ref{22a}) and (\ref{22b}) can be solved numerically by specifying appropriate boundary conditions. Here we follow \cite{b1} and assume that at the center of the configuration the boundary conditions for the gravitational potentials are:
\ben\lb{23}
\widetilde\phi(0)=\widetilde\psi(0)=1,\qquad \frac{d\widetilde\phi}{d\widetilde r}\bigg\vert_{\widetilde r=0}=\frac{d\widetilde\psi}{d\widetilde r}\bigg\vert_{\widetilde r=0}=0.
\een

\section{Analysis of  some fields in the Newtonian and 1PN approximations}

As in the work \cite{b1} we shall analyze in this section the profiles of the mass density,  velocity of test particles in circular motion and  potential energy  as functions of the radial distance.

The mass density can be read from (\ref{20a}) and (\ref{20b}) and  written as a sum of a Newtonian $\widetilde\rho_{\rm N}$ and a post-Newtonian $\widetilde\rho_{\rm PN}$ contribution as
\ben\lb{24}
&&\widetilde \rho=\frac\rho{\rho_0}=\widetilde \rho_{\rm N}+\widetilde \rho_{\rm PN},\qquad\hbox{where}
\\\lb{25a}
&&\widetilde \rho_{\rm N}=e^{\widetilde\phi}{\rm erf}\left(\sqrt{\widetilde\phi}\right)-2\sqrt{\frac{\widetilde\phi}\pi}
\\\no
&&\widetilde \rho_{\rm PN}=\frac1{\zeta_0}\Bigg[\bigg(\frac{3}2+\frac72{\widetilde\phi}-\frac12\widetilde\phi^2+\widetilde\psi\bigg)e^{\widetilde\phi}{\rm erf}\left(\sqrt{\widetilde\phi}\right)
\\\lb{25b}
&&-\bigg(3+\frac{23}2\widetilde\phi-10\widetilde\phi^2+2\widetilde\psi\bigg)\sqrt{\frac{\widetilde\phi}\pi}\Bigg].
\een
In Fig. \ref{Fig1} it is plotted the dimensionless mass density $\widetilde\rho$ as function of the dimensionless radial distance $\widetilde r$ for the Newtonian and post-Newtonian approximations. The post-Newtonian approximation is a function of $\z_0=mc^2/kT_0$, the ratio of the rest energy of the gas particles  and the thermal energy of the gas. One can infer from this figure that the contributions to the mass density becomes larger  at  the configuration center by decreasing the value $\z_0$, i.e., by increasing the temperature of the gas $T_0$. All mass densities tend to zero for large values of the radial distance $\widetilde r$. Opposed to the case where the distribution function is characterized by a polytropic function of the energy \cite{b1}, here none of the mass densities in the 1PN approximation  become negative. However, the solutions for the potentials for values larger than $\widetilde r\approx 3.6$ become complex.

\begin{figure}[h]\vskip0.5cm
\vskip0.7cm
\includegraphics[width=0.45\textwidth]{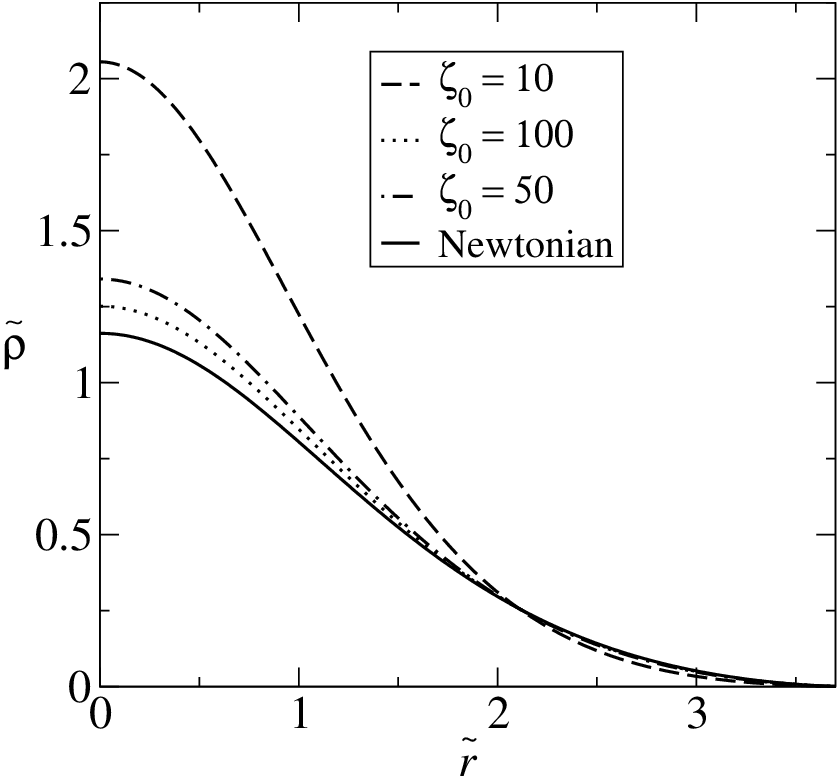}
\caption{Mass density $\widetilde\rho$ as function of the radial distance $\widetilde r$. Newtonian: straight line; Post-Newtonian: dotted line $\z_0=100$, dash-dotted line $\z_0=50$, dashed line $\z_0=10$. }
\lb{Fig1}
\end{figure}

The determination of post-Newtonian corrections to the rotation curves is based on the equation for the acceleration of a free falling particle \cite{Wein,b1}, which for static fields reads
\ben\lb{26}
\frac{d{\bf v}}{dt}=-\nabla\left(\phi+\frac{2\phi^2+\psi}{c^2}\right)+\frac{4\bf v}{c^2}{\bf v}\cdot\nabla\phi-\frac{v^2}{c^2}\nabla\phi.
\een
The radial component of the above equation in spherical coordinates $(r,\theta,\varphi)$ for circular orbits of particles in the equatorial plane where $\dot r=0$, $\dot \theta=0$ and $\theta=\pi/2$ reduces to
\ben\lb{27}
r\dot\varphi^2\left(1-\frac{r}{c^2}\frac{\partial\phi}{\partial r}\right)=\frac{\partial\phi}{\partial r}+\frac{4\phi}{c^2}\frac{\partial\phi}{\partial r}+\frac1{c^2}\frac{\partial\psi}{\partial r}.
\een

The expression for the circular velocity $v_\varphi=r\dot\varphi$ in terms of the gravitational potentials follows from the above equation by considering terms up to the $1/c^2$ order, yielding
\ben\lb{28}
v_\varphi=\sqrt{r\frac{\partial\phi}{\partial r}\left(1+\frac{4\phi}{c^2}+\frac{r}{c^2}\frac{\partial\phi}{\partial r}\right)+\frac{r}{c^2}\frac{\partial\psi}{\partial r}}.
\een
Note that the Newtonian circular velocity is given by $v_\varphi=\sqrt{r\partial\phi/\partial r}$.

Now  in terms of the tilde variables  the circular velocity becomes
\ben\lb{29}
\widetilde v_\varphi=\sqrt{\widetilde r\frac{\partial \widetilde\phi}{\partial \widetilde r}\left(\frac{4\widetilde\phi}{\z_0}+\frac{\widetilde r}{\z_0}\frac{\partial\widetilde\phi}{\partial\widetilde r}-1\right)-\frac{\widetilde r}{\z_0}\frac{\partial\widetilde\psi}{\partial \widetilde r}},
\een
where  $\widetilde v_\varphi=v_\varphi\sqrt{m/kT_0}$ denotes the dimensionless circular velocity.

The  dimensionless circular velocity $\widetilde v_\varphi$ as a function of the dimensionless radial coordinate $\widetilde r$ is plotted in Fig. \ref{Fig2}. The profiles of the circular velocity have the same behaviors for the Newtonian and post-Newtonian approximations, but for the post-Newtonian approximations the circular velocities have large values. Furthermore, as the mass density, the large values of the circular velocity are attained for smaller values of the parameter $\z_0$, which correspond to large values of the temperature $T_0$.  The behavior of the circular velocity  for ideal gases differs from the one where the distribution function is characterized by a polytropic function of the energy \cite{b1}, since in the latter case the values of the circular velocity in the 1PN
approximation are smaller than the corresponding Newtonian approximation. Here the large values of the circular velocity in the 1PN approximation are due to the fact that the increase of the temperature of the gas, increases the thermal velocity of the particles of the gas $\sqrt{kT_0/m}$.

\begin{figure}[h]\
\vskip0.7cm
\includegraphics[width=0.45\textwidth]{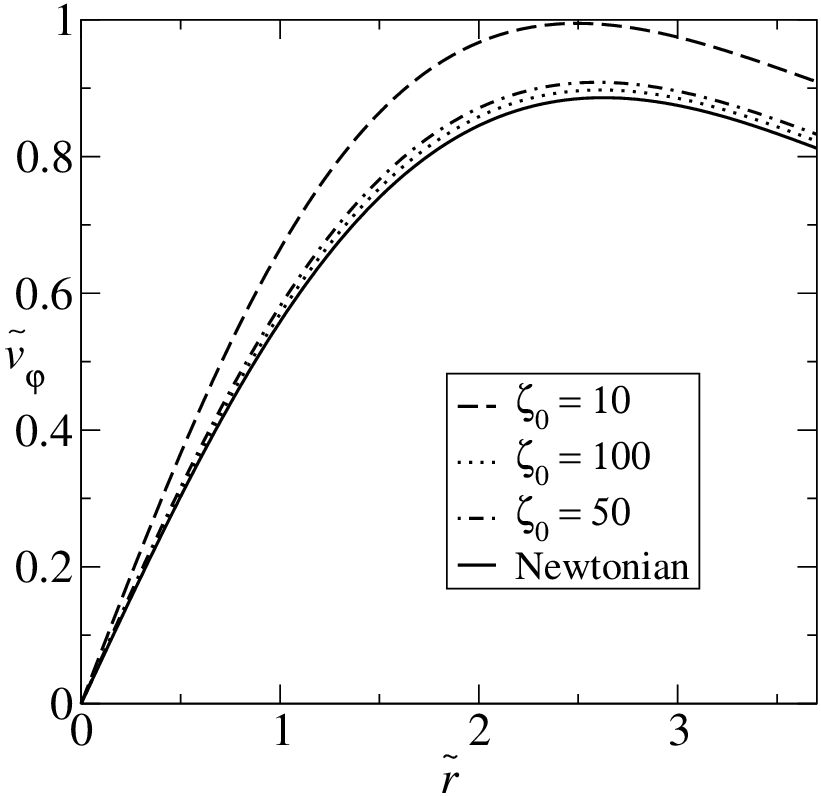}
\caption{ Circular velocity $\widetilde v_\varphi$ as function of the radial distance $\widetilde r$. Newtonian: straight line; Post-Newtonian: dotted line $\z_0=100$, dash-dotted line $\z_0=50$, dashed line $\z_0=10$.}
\lb{Fig2}
\end{figure}

Another field that can be analyzed is the gravitational potential energy of a gas particle, which can be obtained from the expression for the particle energy (\ref{18}) by taking $\bf v=0$. In terms of dimensionless quantities the Newtonian $\widetilde U_{\rm N}$ and post-Newtonian $\widetilde U_{\rm PN}$ gravitational potential energy read
\ben\lb{30a}
\widetilde U=\frac{U}{kT_0}=\widetilde U_{\rm N}+\widetilde U_{\rm PN},\qquad \hbox{where}
\\\lb{30b}
\widetilde U_{\rm N}=-\widetilde\phi,\qquad \widetilde U_{\rm PN}=\frac1{\z_0}\left(\frac{\widetilde\phi^2}2-\widetilde\psi\right).
\een
In Fig. \ref{Fig3} it is shown the graphic representation of the dimensionless gravitational potential energy $\widetilde U$ as function of the dimensionless radial distance $\widetilde r$. We can infer from this figure that the Newtonian gravitational potential energy is always negative, while the post-Newtonian gravitational potential energies change their sign for large values of the radial distance from the configuration center. The temperature of the gas in the post-Newtonian term $\widetilde\phi^2/2\z_0$ determines the sign change of the gravitational potential energy. The gravitational potential energy exhibits the same behavior as the one which takes into account a polytropic function of the energy for the distribution function with  polytropic index $n=3$ \cite{b1}.

\begin{figure}[h]\
\vskip0.7cm
\includegraphics[width=0.45\textwidth]{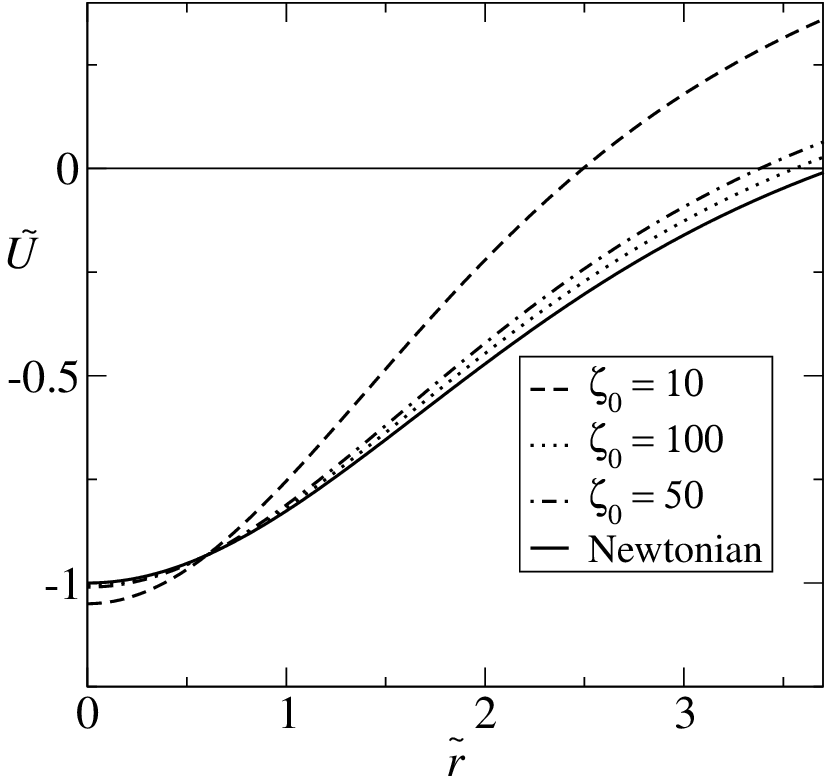}
\caption{Gravitational potential energy $\widetilde U$ as function of the radial distance $\widetilde r$. Newtonian: straight line; Post-Newtonian: dotted line $\z_0=100$, dash-dotted line $\z_0=50$, dashed line $\z_0=10$.}
\lb{Fig3}
\end{figure}

From the energy-momentum tensor given in (\ref{20c}) one can obtain the expression for the pressure of the gas through $p={\buildrel\!\!\!\! _{2} \over{T^{ii}}}/3$. In terms of dimensionless quantities we have
\ben\lb{31}
\widetilde p=\frac{m p}{k\rho_0 T_0}=e^{\widetilde\phi}{\rm erf}\left(\sqrt{\widetilde\phi}\right)-\left(2+\frac43\widetilde\phi\right)\sqrt{\frac{\widetilde\phi}\pi}.
\een
The dimensionless pressure $\widetilde p$ as function of the dimensionless radial coordinate $\widetilde r$
is plotted in Fig. \ref{Fig4}. The behavior of the pressure is the same as the one for the mass density, its maximum
value occurs at the configuration center and it tends to zero for large values of the radial distance. Note that in the 1PN approximation the pressure has no contribution which depends on the factor $1/\z_0$, since these contributions will appear in the order of ${\buildrel\!\!\!\! _{4} \over{T^{ii}}}$.
Fig. \ref{Fig5} shows the behavior of the ratio $p/\rho$ in terms of dimensionless radial coordinate. For small radii, $p/\rho$ goes to a constant value whereas tends to zero at large radii.
\begin{figure}[h]\vskip0.5cm
\vskip0.7cm
\includegraphics[width=0.45\textwidth]{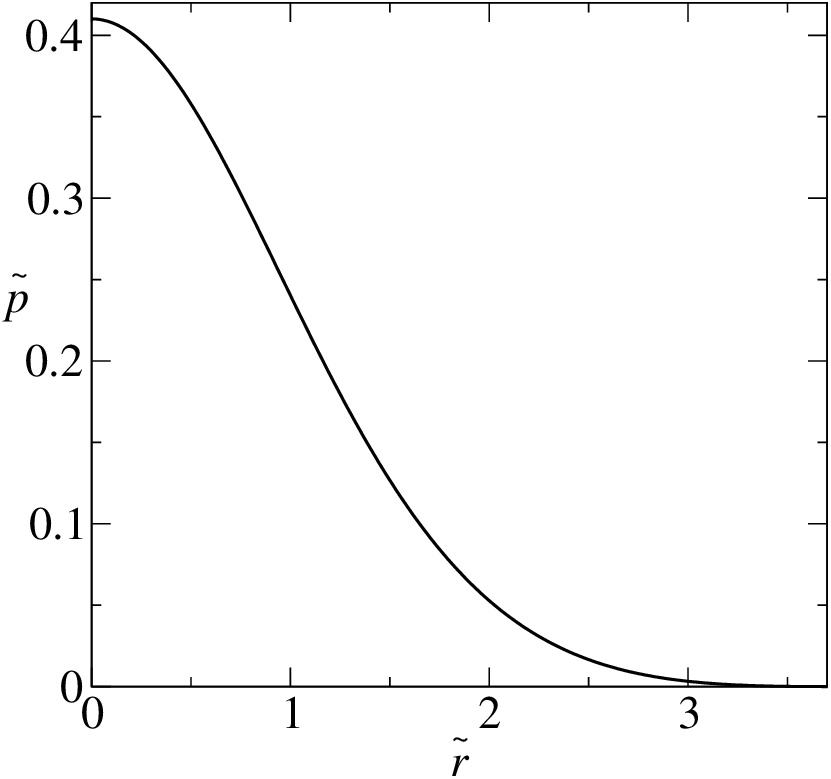}
\caption{Pressure $\widetilde p$ as function of the radial distance $\widetilde r$.}
\lb{Fig4}
\end{figure}

As a final remark we can ask about the influence of the boundary conditions in the behavior of the analyzed fields. It was found that the boundary condition that has more influence on the solutions refers to the Newtonian gravitational potential $\widetilde\phi$ at $\widetilde r=0$. However by considering values of $\widetilde\phi(0)$ in the range [0.5,3] the behavior of the curves does not change, only the absolute values of the fields become smaller or larger than the ones obtained for  $\widetilde\phi(0)=1$.


\begin{figure}[hb]\vskip0.5cm
  \centering
  \includegraphics[width=0.45\textwidth]{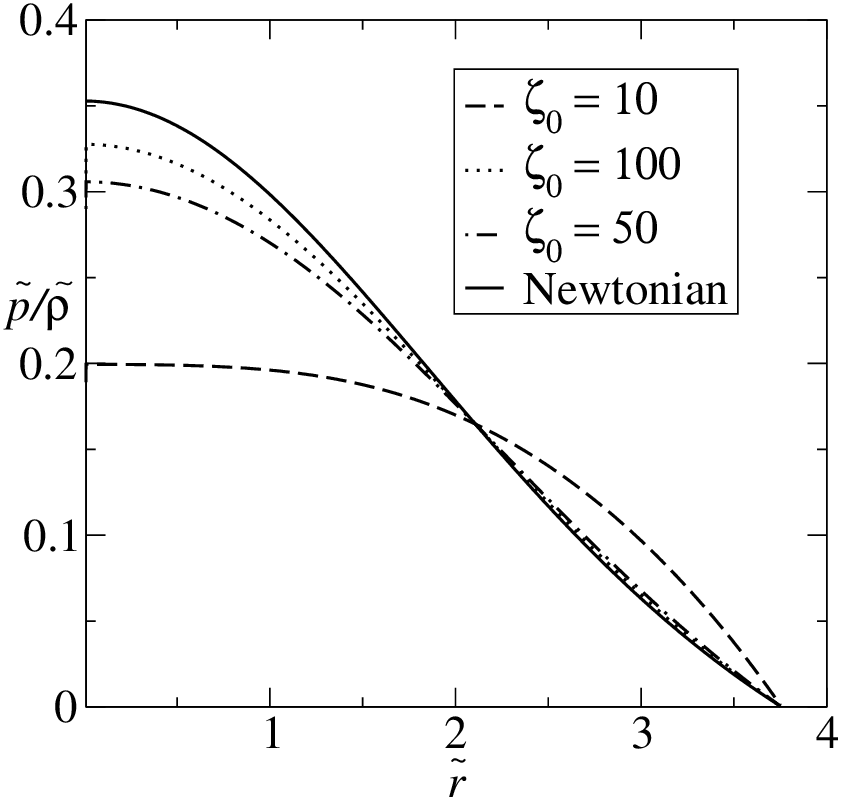}
  \caption
  {Pressure-density ratio  as function of the radial distance $\widetilde r$ for different values of $\z_{0}$}
  \label{Fig5}
\end{figure}



\begin{figure}[hb]\vskip0.5cm
  \centering
  \includegraphics[width=0.45\textwidth]{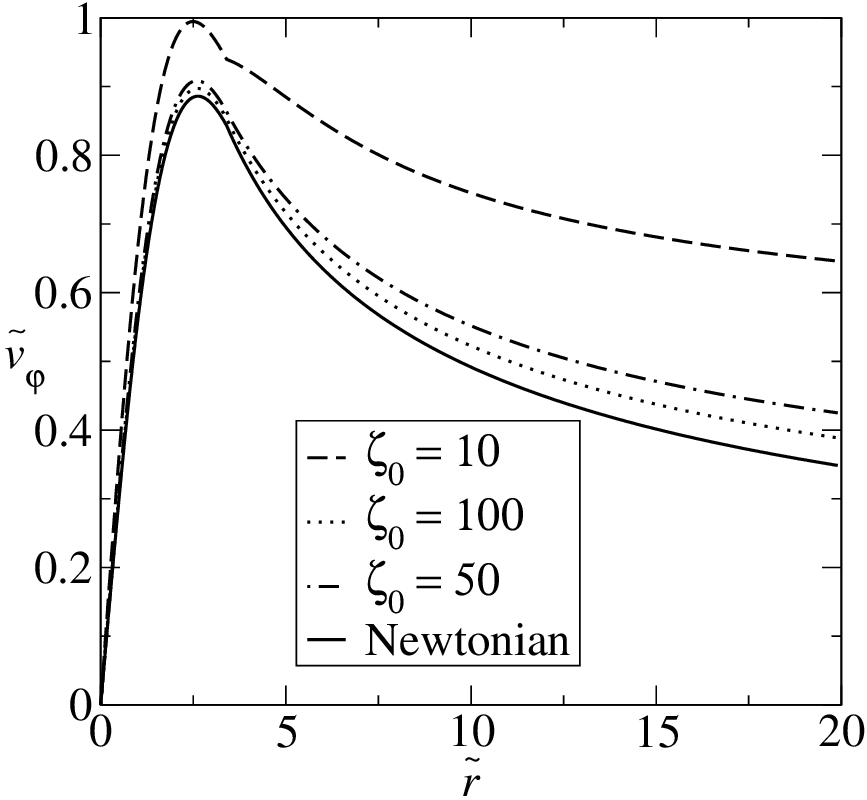}
  \caption
  {Circular velocity rotation curves in terms of the dimensionless radial coordinate for different values of $\z_0$}
  \label{Fig6}
\end{figure}

\section{Circular rotation curve}
As we mentioned in Sec. I, one of the most challenging puzzles  of current astrophysics is
to account for the gravitational mass associated with large structures such as galaxies and cluster or super-cluster of galaxies. It is  well-known that the rotation curves of galaxies do not follow a Newtonian profile, which
indicates that Newtonian gravity fails to predict the mass distribution of such objects. One way to explain such
mismatch is by introducing an unknown component with zero pressure dubbed dark matter. This component cannot be observed or measured
directly but can be detected through
its gravitational effects with the nearby environment (galaxies), in particular, this means that dark matter does not interact directly
with the standard matter (at least not with the ones  contained in  the standard particle model) \cite{ga}.
While in the literature there are  lots of particle-like models for dark matter, none of them seem to be able to
properly describe dark matter completely at galactic scale or cosmological scales, then the only true physical information
about dark matter is how it gravitationally interacts with normal matter (galaxies). Indeed, the gravitational lensing  within a cluster of galaxies, reveals that
the presence of concentrated dark matter in the inner zone of a cluster.  Something that clearly cannot be predicted with ordinary matter within Newtonian theory
because  the mass of the galaxy is related with the circular rotation velocity $V$  through the Newton's law, namely the speed $V$ in a circular orbit at radius $r$ is related to the mass $M(<r)$ interior to that radius by the exact equation $M(<r) = rV^2/G$, being $G$  Newton's constant.
One way to realize about the missing mass fact is by using two well-known procedures. To be more precise,  applying the virial theorem to a cluster,  one can determine the virial mass of the cluster and summing up all the mass of galaxies within the cluster one can estimate the baryonic mass. It turns out  that the former one is 20 times bigger than the latter one, showing in this way the existence of additional (nonbaryonic) matter in the cluster \cite{ga}.

So far we have focused on the basic features of self-gravitating systems of ideal gases in the 1PN approximation. Now, we devote our attention to the circular
rotation curve due to its physical relevance in modern astrophysics at theoretical and observational levels, and in particular we examine the matching of two
solutions at a critical radius, say $\widetilde{r}_c$. The idea behind such standard procedure is to reproduce the well-known observational features of the galaxy rotation curves, namely,  a linear regime at small radii, passes through a cusp,  and ends with a flatten shape at large radii. Let us now show how this shape can be obtained with the model at hand within the 1PN approximation.  Fig. \ref{Fig2} shows that model describes well the inner zone corresponding to $\widetilde{r} \leq \widetilde{r}_c$ because the circular rotation curve grows linear with the radial distance,  then it passes through small cusp  and becomes ill-defined for $\widetilde{r} >\widetilde{r}_c + \epsilon$ with $\epsilon>0$, implying that the gravitational potentials become complex and therefore the solutions are  unphysical. In order to solve this issue, we match these solutions with another  two gravitational potentials, called $\widetilde{\phi}_{2}(\widetilde{r})$ and $\widetilde{\psi}_{2}(\widetilde
 {r})$. A somewhat natural question to ask is this one: which kind of physical equation should satisfy these new potentials?  A quick answer would be that these ones must fulfill the Laplace equation, namely $\nabla^{2} \widetilde{\phi}_{2}(\widetilde{r})=0$ and $\nabla^{2}\widetilde{\psi}_{2}(\widetilde{r})=0$, and the potentials and theirs first derivatives must be glued at $\widetilde{r}_c$  in order to have a well-defined boundary problem. Let us analyze this proposal in detail. Taking into account that both potentials satisfy the Laplace equation, they can be recast as $\widetilde{\phi}_{2}(\tilde{r})=\alpha/\tilde{r} + \beta$  and  $\widetilde{\psi}_{2}(\tilde{r})=\gamma/\tilde{r} + \delta$. After imposing the continuity of the potential and its derivative at $\widetilde{r}_c$ , we arrived at $\alpha=-\widetilde{r}^2_c \widetilde{\phi'}_{1}(\widetilde{r}_c)$ along with  $\beta=\widetilde{\phi}_{1}(\widetilde{r}_c)+\widetilde{r}_c\widetilde{\phi}'_{1}(\widetilde{r}_c)$. The same thing holds for the other potential by replacing $\alpha \rightarrow \gamma$,  $\beta \rightarrow \delta$ along with $\widetilde{\phi}_{1} \rightarrow \widetilde{\psi}_{1}$, where $\widetilde{\phi}_{1}$  and $\widetilde{\psi}_{1}$ are the gravitational potentials which satisfy the Poisson equations (\ref{22a})-(\ref{22b}) in the inner zone ($\widetilde{r} \leq \widetilde{r}_c$).  With this proposal at hand, we can extend physically our first solution beyond the critical radius, $\widetilde{r}_c$, however, we can prove that the aforesaid gravitational potentials do not reproduce the desired property of a flatten circular rotation curve in the outer zone. Replacing   both potentials $\widetilde{\phi}_{2}(\widetilde{r})$ and $\widetilde{\psi}_{2}(\widetilde{r})$ into (\ref{29}), one can check that the circular rotation curve  goes to zero  in the limit of  large radii. Therefore, we need to introduce another type of potentials to do the job.  In order to do so, we propose that  $\widetilde
  {\phi}_{2}(\tilde{r})$  satisfies the Laplace's equation whereas $\widetilde {\psi}_{2}(\tilde{r})$ fulfills a Poisson's equation with the same boundary condition mentioned above. The latter potential can be parametrized as
\ben\lb{32}
\widetilde {\psi}_{2}(\tilde{r})= \gamma \frac{e^{-k \widetilde{r}}}{\widetilde{r}}+ \delta \ln \widetilde{r}.
\een
Here $k$ is a parameter of the Yukawa term while   $\gamma$ and $\delta$ are integration constants. The latter ones are obtained by solving the following linear system
\begin{equation}
\left[\begin{array}{c} \widetilde {\psi}_{2}(\tilde{r}_c) \\ \widetilde {\psi}'_{2}(\tilde{r}_c)\end{array} \right] = \begin{bmatrix}\frac{e^{- k\widetilde{r}_c}}{\widetilde{r}_c} &\ln \widetilde{r}_c  \\ -k\frac{e^{- k\widetilde{r}_c}}{\widetilde{r}_c} -\frac{e^{- k\widetilde{r}_c}}{\widetilde{r}^2_c}& \frac{1}{\widetilde{r}_c} \end{bmatrix} \times \left[ \begin{array}{c} \gamma \\ \delta \end{array} \right].
\end{equation}
Using $\widetilde {\psi}_{2}(\tilde{r}_c)=\widetilde {\psi}_{1}(\tilde{r}_c)$ and $\widetilde {\psi}'_{2}(\tilde{r}_c)=\widetilde {\psi}'_{1}(\tilde{r}_c)$ we find the value of the integration constants numerically at $\tilde{r}_c \simeq 3.4001$, namely $\gamma \simeq -29.7253$ $\delta \simeq -2.6906$. It should be pointed that the aforesaid procedure is not dependent of the value taken by $\z_{0}$, however, the circular rotation curve is. As can be seen from Fig. \ref{Fig6}, we find that the circular rotation curves flatten at very large radii for different values of $\z_{0}$. It is easy to understand how this proposal reproduces the flat shape by replacing  the Coulomb potential $\widetilde {\phi}_{2} (\tilde{r})$ along with $\widetilde {\psi}_{2} (\tilde{r})$ into   (\ref{29}). The Coulomb and Yukawa terms  fade away for large radii but the logarithm contribution introduces a constant in the  circular velocity which dominates for large radii. In this way, we could  include a Coulomb term in $\widetilde {\psi}_{2} (\tilde{r})$ instead of the Yukawa term and it will also lead us to the right flatten curve, however, the Yukawa term is better because it behaves  smoother than the Coulomb potential for large radii.
We end this section by mentioning some physical outcomes of our previous proposal. For $\widetilde{r} \geq \widetilde{r}_c$  the Newtonian density  is given
by $\widetilde{\rho}_{mass}=-\nabla^{2}\widetilde {\phi}_{2}=\alpha/{\widetilde{r}}^2$ which clearly goes to zero for large radius.  In this zone, the post-Newtonian density and pressure are entangled provided we only know $\widetilde {\psi}_{2}$, that is, $\widetilde{\rho}_{PN}+\widetilde{p}_{PN} =-\nabla^{2}\widetilde {\psi}_{2}=-\gamma e^{-\widetilde{r}}/\widetilde{r}-\delta/{\widetilde{r}}^2$, where we have taken   $k=1$ without loss of generality, then $\widetilde{\rho}_{PN}+\widetilde{p}_{PN}$ also vanishes in this limit. Notice that $\widetilde{\rho}_{PN}+\widetilde{p}_{PN}$ is always positive  because $\gamma<0$ and $\delta<0$. All in all, we showed that rotation curves with a flatten profile can be obtained within the post-Newtonian approximation, starting from the Maxwell-J\"uttner distribution function  at 1PN and solving the corresponding gravitational potential at 1PN order.  As a final remark, we should emphasize that in our approach the ergodic function used differs
  substantially from the polytrope profile explored in \cite{b1}.
\section{Conclusions}
We have built an astrophysical model based on the Maxwell-J\"uttner distribution function   within the framework of general relativity which is described using post-Newtonian formalism. In order to obtain this expression, we have integrated over the peculiar velocity 3D space, keeping only the relevant terms up to 1PN order.  With this ergodic distribution at hand, we have obtained the general form of Newtonian density along with the  post-Newtonian density and pressure terms which enter in the energy-momentum tensor by using several Gaussian integrals (cf. Appendix B). This allowed us to demonstrate that our general expression coincides with the one reported by Weinberg \cite{Wein}, validating our procedure. As an example of how to apply this procedure, we considered the case of particle four-flow and calculated  its temporal and spatial components.

From the energy density and pressure terms at 1PN, we looked for static solution by analyzing the boundary value problem. We found that  the energy density, pressure  and gravitational potentials  profiles in terms of dimensionless radial coordinate by solving the aforesaid equations numerically.  In particular, we found the energy density vanishes for large radii but approaches to a constant value at the origin, further,  such value becomes larger for smaller values of $\z_0$ [see Fig. \ref{Fig1}]. Contrary to the case of polytropic ergodic distribution explored in \cite{b1}, the energy density remains always positive  at 1PN order. As part of the process of evaluating  the circular  velocity profiles, we obtained that post-Newtonian curves reach larger values in relation with the Newtonian case. In both cases, we have found that these curves exhibit a linear behavior near the center and then pass through a cusp [see Fig. \ref{Fig2}]. Interestingly enough, we have found
 the behavior of the circular velocity  for ideal gases differs from the one where the distribution function is characterized by a polytropic function of the energy \cite{b1}, since in the latter case the values of the circular velocity in the 1PN approximation are smaller than the corresponding Newtonian approximation. In our case,  the large values of the circular velocity in the 1PN approximation are due to the fact that the increase of the temperature of the gas, increases the thermal velocity of the particles of the gas $\sqrt{kT_0/m}$. We  have examined the behavior of the dimensionless gravitational potential energy $\widetilde U$. So we have found that the Newtonian gravitational potential energy is always negative, while the post-Newtonian gravitational potential energies change their sign for large values of the radial distance from the configuration center [see Fig. \ref{Fig3}]. The temperature of the gas in the post-Newtonian term $\widetilde\phi^2/2\z_0$ determines the sign change of the gravitational potential energy. The gravitational potential energy exhibits the same behavior that the one associated with a polytropic distribution function with polytropic index $n=3$. In addition,  we found a parametric profile of the equation of state $p(\rho)$ in terms of the dimensionless radial coordinate, see Fig. \ref{Fig5}. At small radii, the ratio pressure-density becomes almost constant; increasing $\z_{0}$ the constant reaches larger values or equivalently  the thermal energy of the ideal gas decreases considerably. The situation is reversed at large radii.

We have patched together two different kinds of gravitational potentials at a critical radius, called ${\widetilde r}_c$, in order to have a well-defined boundary problem, provided the potentials became complex beyond this radius.  The physical motivation in dealing with this issue was to select gravitational potentials which reproduce a flatten behavior  in the rotation curve at large radii, provided the linear and the cusp zones were already described by those potentials obtained by integrating numerically the energy density and pressure coming from the Maxwell-J\"uttner distribution function at 1PN. In doing so,  we have shown that one of the potentials has a Coulomb form while the other one presents two kinds of terms, a Yukawa contribution plus a logarithmic term. Indeed, the term responsible for the constant circular velocity at large radii turned out to be the logarithmic term. Regarding the changes introduced by these solutions, we have found that the Coulomb potential
 led to an inverse square power law for the Newtonian density, as one could expect. However,  we could not disentangle the post-Newtonian density from the post-Newtonian pressure, thus we found   $\widetilde{\rho}_{PN}+\widetilde{p}_{PN} =-\gamma e^{-\widetilde{r}}/\widetilde{r}-\delta/{\widetilde{r}}^2$ which leads to  a positive quantity, vanishing for large radii only. 
\vspace{1.cm}
\acknowledgments
G. M. K. and M. G. R. are supported by Conselho Nacional de Desenvolvimento Cient\'ifico e Tecnol\'ogico (CNPq)  and K. W. by Coordena\c c\~ao ̧de Aperfei\c coamento
de Pessoal de N\'ivel Superior (CAPES). G. M. K. thanks J. F. Pedraza for suggestions.

\section*{Appendix A: SEVERAL QUANTITIES AT 1PN}

Up to the 1PN fourth order  the components of the metric tensor are given by \cite{Wein}
\ben\lb{a1}
&&g_{00}=-1+{\buildrel\!\!\!\! _{2} \over{g_{00}}}+{\buildrel\!\!\!\! _{4} \over{g_{00}}}+\dots
\\\lb{a2}
&&g_{ij}=\delta_{ij}+{\buildrel\!\!\!\! _{2} \over{g_{ij}}}+{\buildrel\!\!\!\! _{4} \over{g_{ij}}}+\dots
\\\lb{a3}
&&g_{0i}={\buildrel\!\!\!\! _{3} \over{g_{0i}}}+{\buildrel\!\!\!\! _{5} \over{g_{0i}}}+\dots
\\\lb{a4}
&&{\buildrel\!\!\!\! _{2} \over{g_{00}}}=-2\frac{\phi}{c^2},
\qquad
{\buildrel\!\!\!\! _{4} \over{g_{00}}}=-\frac2{c^4}\left(\phi^2+\psi\right),
\\\lb{a5}
&&{\buildrel\!\!\!\! _{2} \over{g_{ij}}}=-2\delta_{ij}\frac{\phi}{c^2},\qquad
{\buildrel\!\!\!\! _{3} \over{g_{0i}}}=\frac{\xi_i}{c^3},
\\\lb{a6}
&&\sqrt{-g}=1-\frac{2\phi}{c^2}-\frac{\phi^2-\psi}{c^4}.
\een
 Here $\phi$ is the Newtonian gravitational potential, and $\psi$ and $\xi_i$ are gravitational  potentials in the 1PN approximation. These gravitational potentials are connected with the energy-momentum tensor by Einstein's field equations.

\section*{Appendix B: GAUSSIAN INTEGRALS}
Let us summarize the most interesting cases that we have employed within this article. Just to make things more familiar, let us consider integration over a 3D space associated to the peculiar velocity space defined as $W*=\{(w^{*}, \theta, \phi): 0 \leq w^{*}\leq \infty, 0 \leq \theta \leq 2\pi, 0 \leq \phi\leq \pi \}$, where $w^{*}$ is a dimensionless peculiar velocity and the volume element is given by $d^{3}W^{*}= {w^{*}}^{2}d w^{*} d\Omega_{2}$ with $d\Omega_{2}$ the 2-dimensional element of solid angle. Let us consider the Gaussian distribution $F(w^{*})=e^{-{w^{*}}^{2}/2}$ so the following expressions can be obtained:
\ben\lb{c1}
&&I_{0}=\int{e^{-{\frac{{w^{*}}^{2}}{2} }} d^{3}W^{*}}= 4\pi\sqrt{\frac{\pi}{2}},
\\\lb{c2}
&&I_{2}=\int{{w^{*}}^{2}e^{-{\frac{{w^{*}}^{2}}{2} }} d^{3}W^{*}}=12\pi \sqrt{\frac{\pi}{2}},
\\\lb{c3}
&&I_{4}=\int{{w^{*}}^{4}e^{-\frac{{w^{*}}^{2}} {2}} d^{3}W^{*}}=60\pi \sqrt{\frac{\pi}{2}},
\\\lb{c4}
&&I_{6}=\int{{w^{*}}^{6}e^{-{\frac{{w^{*}}^{2}}{2} }} d^{3}W^{*}}=420\pi \sqrt{\frac{\pi}{2}}.
\een
Using (\ref{c2}), (\ref{c3}), and (\ref{c4})  we can prove that the following expressions hold (see e.g. \cite{GK2}):
\ben\lb{d1}
&&\int{F(w^{*})w^{*}_{i}w^{*}_{j} d^{3}W^{*}}=\frac{I_{2}}{3} \delta_{ij},
\\\no
&& \int{F(w^{*})w^{*}_{i}w^{*}_{j}w^{*}_{k} w^{*}_{l}  d^{3}W^{*}} \\\lb{d2}
&&=\frac{I_{4}}{15}[\delta_{ij}\delta_{kl}+ \delta_{ik}\delta_{jl}+\delta_{il}\delta_{jk}] ,
\\\no
&&\int{F(w^{*})w^{*}_{i}w^{*}_{j}w^{*}_{k} w^{*}_{l} w^{*}_{m} w^{*}_{n}  d^{3}W^{*}}
\\\no
&&=\frac{I_{6}}{105}[\delta_{ij}\delta_{kl}\delta_{mn} + \delta_{ij}\delta_{km}\delta_{ln}+\delta_{ij}\delta_{kn}\delta_{lm}
\\\no
&&+ \delta_{ik}\delta_{jl}\delta_{mn} +\delta_{ik}\delta_{jm}\delta_{ln} + \delta_{ik}\delta_{jn}\delta_{lm}+\delta_{il}\delta_{jk}\delta_{mn}
\\\no
&&+ \delta_{il}\delta_{jm}\delta_{kn} +\delta_{il}\delta_{jn}\delta_{km} + \delta_{im}\delta_{jk}\delta_{ln}+\delta_{im}\delta_{jl}\delta_{kn}
\\\lb{d3}
&&+ \delta_{im}\delta_{jn}\delta_{kl}+\delta_{in}\delta_{jk}\delta_{lm}+ \delta_{in}\delta_{jl}\delta_{km}+\delta_{in}\delta_{jm}\delta_{kl} ].\qquad
\een

\end{document}